\documentclass[preprint,12pt]{elsarticle}
\usepackage{bm,amsmath,amssymb}
\usepackage{graphicx}
\usepackage{subfigure}
\usepackage{color}
\usepackage{soul}
\usepackage{hyperref}
\usepackage{multirow}
\usepackage{bm,amsmath,amssymb}
\usepackage{graphicx}
\usepackage{subfigure}
\usepackage{color}
\usepackage{soul}
\usepackage{hyperref}
\usepackage{multirow}
\usepackage{lineno}
\setulcolor{blue}
\setstcolor{red}
\sethlcolor{yellow}

\newcommand \etal {{\sl et al.}}

\newcommand \eq[1] {Eq.\ (\ref{#1})}
\newcommand \beq {\begin{equation}}
\newcommand \eeq {\end{equation}}
\newcommand \beqa {\begin{eqnarray}}
\newcommand \eeqa {\end{eqnarray}}

\newcommand \hmu {\hat{\mu}}
\begin{document}

\title{The melting and abundance of open charm hadrons}

\author[1]{A. Bazavov}
\author[2]{H.-T. Ding}
\author[2]{P. Hegde} 
\author[3]{O. Kaczmarek}
\author[3,4]{F. Karsch}
\author[3]{\\ E. Laermann}
\author[3]{Y. Maezawa}
\author[4]{Swagato Mukherjee}
\author[4,5]{H. Ohno}
\author[4]{\\ P. Petreczky}
\author[3]{C. Schmidt}
\author[3]{S. Sharma\corref{cor1}}
\ead{sayantan@physik.uni-bielefeld.de}
\cortext[cor1]{Corresponding author}
\author[6]{W. Soeldner}
\author[7]{M. Wagner}

\address[1]{Department of Physics and Astronomy, University of Iowa, Iowa City, IA
52240, USA}
\address[2]{Key Laboratory of Quark \& Lepton Physics(MOE) and 
Institute of Particle Physics, Central China Normal University, 
Wuhan, 430079, China}
\address[3]{Fakult\"at f\"ur Physik, Universit\"at Bielefeld, D-33615 Bielefeld,
Germany}
\address[4]{Physics Department, Brookhaven National Laboratory, Upton, NY 11973, USA}
\address[5]{Center for Computational Sciences, University of Tsukuba, Tsukuba,
Ibaraki 305-8577, Japan}
\address[6]{Institut f\"ur Theoretische Physik, Universit\"at Regensburg, D-93040
Regensburg, Germany}
\address[7]{Physics Department, Indiana University, Bloomington, IN 47405, USA}

\date{\today}
\begin{abstract}
\small
Ratios of cumulants of conserved net charge fluctuations are sensitive to the degrees
of freedom that are carriers of the corresponding quantum numbers in different phases
of strong interaction matter.  Using lattice QCD with 2+1 dynamical flavors and quenched 
charm quarks we calculate second and fourth order cumulants of net charm fluctuations and 
their correlations with other conserved charges such as net baryon number, electric
charge and strangeness. Analyzing appropriate ratios of these cumulants we probe the nature
of charmed degrees of freedom in the vicinity of the QCD chiral crossover region.
We show that for temperatures above the chiral crossover transition temperature, charmed degrees of freedom 
can no longer be described by an uncorrelated gas of hadrons. This suggests that the dissociation of 
open charm hadrons and the emergence of deconfined charm states sets in just near the chiral crossover 
transition. Till the crossover region we compare these lattice QCD results with two 
hadron resonance gas models --including only the experimentally established charmed resonances 
and also including additional states predicted by quark model and lattice QCD calculations. 
This comparison provides evidence for so far unobserved charmed hadrons that contribute to the 
thermodynamics in the crossover region.

\end{abstract}


\maketitle

\newpage

\section{Introduction}

Bound states of heavy quarks, in particular the charmonium state $J/\psi$ and its
excitations as well as the heavier bottomonium states, are sensitive probes for
deconfining features of the quark-gluon plasma (QGP) \cite{matsui}. Different
excitations of these states are expected to dissolve at different temperatures in the
QGP, giving rise to a characteristic sequential melting pattern \cite{mehr}. Recent
lattice QCD calculations of thermal hadron correlation functions suggest that certain
quarkonium states survive as bound states in the QGP well beyond the pseudo-critical
temperature of the chiral crossover transition $T_c=(154\pm 9)$~MeV \cite{bazavov};
the $J/\psi$ and its pseudo-scalar partner $\eta_c$ disappear at about $1.5 T_c$
\cite{ding}, while the heavier bottomonium ground states can survive even up to $2
T_c$ \cite{Petreczky,aarts}. 

Light quark bound states, on the other hand, dissolve already at or close to the
pseudo-critical temperature, $T_c$, reflecting the close relation between the chiral
crossover and deconfinement of light quark degrees of freedom. This leads to a sudden
change in the bulk thermodynamic observables and is even more apparent in the
behavior of fluctuations of conserved charges, i.e. baryon number, electric charge or
strangeness \cite{Koch, Ejiri}.  The sudden change of ratios of different moments
(cumulants) of net-charge fluctuations and their correlations in the transition
region directly reflects the change of degrees of freedom that carry the relevant
conserved charges. The total number of hadronic degrees of freedom, i.e. the detailed
hadronic mass spectrum also influences bulk thermodynamics. For instance, the strong
rise of the trace anomaly $(\epsilon -3P)/T^4$, found in lattice QCD calculations may
be indicative for contributions of yet unobserved hadron resonances \cite{Majumder}.

Recently it has been shown that the large set of fourth order cumulants of charge
fluctuations and cross-correlations among fluctuations of conserved charges allows
for a detailed analysis of the change from hadronic to partonic degrees of freedom in
different charge sectors \cite{strange}.  For instance, changes of degrees of freedom
in the strange meson and baryon sectors of hadronic matter can be analyzed separately
by choosing appropriate combinations of charge fluctuation observables. This led to
the conclusion that a description of strong interaction matter in terms of
uncorrelated  hadronic degrees of freedom breaks down for all strange hadrons in the
chiral crossover region, i.e.  at $T\lesssim160$~MeV \cite{strange}, which suggests
that strangeness gets dissolved at or close to $T_c$. This finding has been confirmed
with the analysis presented in \cite{Bellwied}.

A more intriguing question is what happens to the charmed sector of the hadronic
medium at the QCD transition temperature. While it seems to be established that
charmonium states, i.e. bound states with hidden charm, still exist in the QGP at
temperatures well above $T_c$, this may not be the case for heavy-light mesons or
baryons, i.e. open charm mesons ($D$, $D_s$) \cite{rapp,tolos} or charmed baryons
($\Lambda_c,\ \Sigma_c,\ \Xi_c,\ \Omega_c$).  To address this question we calculate
cumulants of net-charm fluctuations as well as correlations between moments of
net-charm fluctuations and moments of net baryon number, electric charge or
strangeness fluctuations. Motivated by the approach outlined in Ref.~\cite{strange}
we analyze ratios of observables that may, at low temperature, be interpreted as
contributions of open charm hadrons to the partial mesonic or baryonic pressure of
strong interaction matter.  We show that a description of net charm fluctuations in
terms of models of uncorrelated hadrons breaks down at temperatures close to the
chiral crossover temperature.  We furthermore show that at low temperatures the
partial pressure calculated in the open charm sector is larger than expected from
hadron resonance gas (HRG) model calculations based on all experimentally measured
charmed resonances as listed in the particle data tables \cite{PDG}. It, however,
agrees well with an HRG based on charm resonances from quark model \cite{Isgur,cQM,Ebertm,Ebert}
and lattice QCD calculations \cite{Prelovsek,Moir,Edwards}.  This points at the
existence and thermodynamic importance of additional, experimentally so far not
established, open charm hadrons.

\section{The charmed hadron resonance gas}

While light quark fluctuations can be quite well described by a hadron resonance gas
\cite{hotQCDHRG} built up from experimentally measured resonances that are listed in
the particle data tables \cite{PDG} it is not at all obvious that this suffices in
the case of the heavy open charm resonances. The particle data tables only list a few
measured open charm resonances. Many more are predicted in the relativistic quark
model \cite{Isgur,cQM,Ebertm,Ebert} and lattice QCD \cite{Moir,Edwards} calculations.
In fact, the large set of excited charmed mesons and baryons found in lattice QCD
calculations closely resembles the excitation spectrum predicted in quark model
calculations.  It is expected that many new open flavor states will be detected in
upcoming experiments at Jefferson Laboratory, FAIR and the LHC
\cite{cQM,glueX,PANDA,LHCb}.  If these resonances are indeed part of the charmed
hadron spectrum of QCD, they become excited thermally and contribute to the
thermodynamics of the charmed sector of a hadron resonance gas. They will show up as
intermediate states in the hadronization process of a quark-gluon plasma formed in
heavy ion collisions and influence the abundances of various particle species
\cite{PBM}. Heavy-light bound states also play an important role in the break-up of
quarkonium bound states. In lattice QCD calculations their contribution becomes
visible in the analysis of the heavy quark potential where they can help to explain
the non-vanishing expectation value of the Polyakov loop at low temperatures
\cite{Megias,Peter}.

In order to explore the significance of a potentially large additional set of open
charm resonances in thermodynamic calculations at low temperature we have constructed
HRG models based on different sets of open charm resonances. In addition to the HRG
model that is based on all experimentally observed charmed hadrons (PDG-HRG), we also
construct an HRG model based on a set of charmed hadrons calculated in a quark
model (QM-HRG) where we used the charmed meson \cite{Ebertm} and charmed
baryon \cite{Ebert} spectrum calculated by Ebert \etal \footnote{The thermodynamic
considerations presented here are mainly sensitive to the number of additional
hadrons included in the calculations and not to the precise values of their masses.
Thus lattice QCD results on the charmed baryon spectra \cite{Edwards} also lead to
similar conclusions.}. 

One may wonder whether all the resonances calculated in a quark model exist or are
stable and long-lived enough to contribute to e.g. the pressure of charmed hadrons.
However, as highly excited states with masses much larger than the ground state
energy in a given quark flavor channel are strongly Boltzmann suppressed, they
play no significant role in thermodynamics.  For this reason we also need not
consider multiple charmed baryons or open charm hybrid states that have been
identified in lattice QCD calculations \cite{Moir,Edwards} but generally have masses
more than (0.8-1)~GeV above those of the ground state resonances.  We explore the
impact of such heavy states by introducing different cut-offs to the maximum mass up to which open charm
resonances are taken into account in the HRG model.  For instance, QM-HRG-3 includes
all charmed hadron resonances determined in quark model calculations that have masses
less than $3$~GeV.

We calculate the open charm meson ($M_C(T,\vec{\mu})$) and baryon
($B_C(T,\vec{\mu}))$ pressure in units of $T^4$, such that the total charm contribution to
the pressure is written as $P_C(T,\vec{\mu})/T^4 = M_C(T,\vec{\mu}) +
B_C(T,\vec{\mu})$. As the charmed states are all heavy compared to the scale of the
temperatures relevant for the discussion of the thermodynamics in the vicinity of the
QCD crossover transition, a Boltzmann approximation is appropriate for all charmed hadrons, 
\begin{eqnarray}
M_C(T,\vec{\mu})\hspace*{-0.3cm} &=&\hspace*{-0.3cm} {1\over {2\pi^2}}\hspace*{-0.3cm} 
\sum_{i\in C-mesons} \hspace*{-0.4cm}
g_i \left(\frac{m_i}{T}\right)^2 K_2({{m_i/T}}) \cosh \left( 
Q_i \hat{\mu}_Q + S_i\hat{\mu}_S + C_i \hat{\mu}_C \right)    \; , \\
B_C(T,\vec{\mu})\hspace*{-0.3cm} &=&\hspace*{-0.3cm}  {1\over {2\pi^2}} \hspace*{-0.3cm}
\sum_{i\in C-baryons} \hspace*{-0.5cm}
g_i \left(\frac{m_i}{T}\right)^2 K_2({{m_i/T}}) \cosh \left( 
B_i\hat{\mu}_B
+ Q_i \hat{\mu}_Q +S_i\hat{\mu}_S+ C_i \hat{\mu}_C \right)   \ . \nonumber
\label{Cpressure}
\end{eqnarray}
Here, $\vec{\mu}=(\mu_B, \mu_Q, \mu_S, \mu_C)$, $\hat{\mu}\equiv \mu/T$ and $g_i$ are
the degeneracy factors for the different states with electric charge $Q_i$,
strangeness $S_i$ and charm $C_i$.

\begin{figure}[!th]
\begin{center}
\includegraphics[scale=0.6]{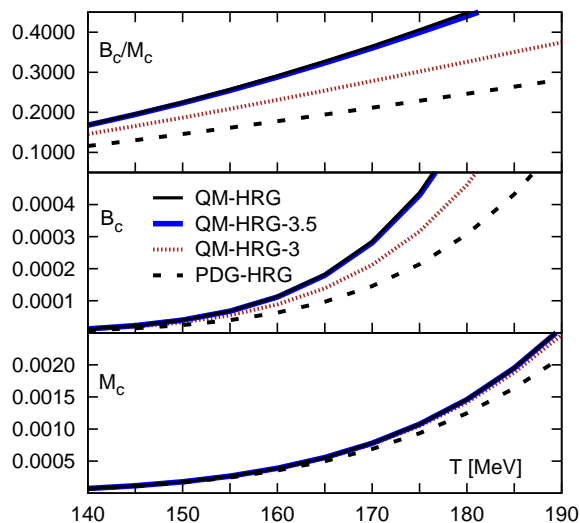}
\end{center}

\caption{Partial pressure of open charm mesons ($M_c$, bottom), baryons ($B_c$,
middle) and the ratio $B_C/M_C$ (top) in a gas of uncorrelated hadrons, using all 
open charm resonances listed in the particle data table (PDG-HRG, dashed lines) \cite{PDG} 
and using additional charm resonances calculated in a relativistic quark model (QM-HRG, solid lines) \cite{Ebertm,
Ebert}. Also shown are results from HRG model calculations where the open charm
resonance spectrum is cut off at mass 3~GeV (QM-HRG-3) and 3.5~GeV (QM-HRG-3.5). At
temperatures below 160~MeV the latter coincides with the complete QM-HRG model results
to better than 1~\%.  }

\label{fig:hadronsPDG}
\end{figure}

Results from calculations of open charm meson and baryon pressures using different
HRG models are shown in Fig.~\ref{fig:hadronsPDG}.  The influence of additional
states predicted by the quark model is clearly visible already in the QCD crossover transition
region. At $T_c$, differences between PDG-HRG
(dashed lines) and QM-HRG (solid lines) in the baryon sector are as large as 40\%
while they are negligible in the meson sector.  This reflects that the experimentally
known meson spectrum is more complete than the baryon spectrum. 

In the open charm meson sector, the well established excitations cover a mass range of
about $700$~MeV above the ground state $D,\ D_s$-mesons.  In the charmed baryon
sector much less is known, for instance, experimentally well known excitations of $\Xi_c$ range up to 
$350$~MeV above the ground state and in the doubly strange charmed baryon sector only 
two $\Omega_c$ states separated by $100$~MeV are well established. 

As a consequence of the limited knowledge of the charmed baryon spectrum compared
to the open charm meson spectrum, the ratio of partial pressures in the baryon and
meson sectors differs strongly between the PDG-HRG and the QM-HRG. This is shown in
Fig.~\ref{fig:hadronsPDG}~(top). Significant differences between the QM-HRG-3 and 
PDG-HRG results also indicate that almost half of the enhanced contributions actually 
comes from additional charmed baryons that are lighter than the heaviest PDG state.
Similar conclusions can be drawn when analyzing partial pressures in the strange-charmed 
hadron sector or the electrically charged charmed hadron sectors.  

\section{Calculation of charm fluctuations in (2+1)-flavor lattice QCD}

In order to detect changes in the relevant degrees of freedom that are the carriers
of charm quantum numbers at low and high temperatures as well as to study their
properties we calculate dimensionless generalized susceptibilities of conserved
charges,
\beq
\chi_{klmn}^{BQSC} = \left. 
\frac{\partial^{(k+l+m+n)} [P(\hmu_B,\hmu_Q,\hmu_S,\hmu_C)/T^4]}
{\partial \hmu_B^k \partial \hmu_Q^l \hmu_S^m \partial \hmu_C^n} 
\right|_{\vec{\mu}=0}
\ .
\label{eq:susc}
\eeq
Here $P$ denotes the total pressure of the system. In the following we also use the
convention to drop a superscript in $\chi_{klmn}^{BQSC}$ when the corresponding
subscript is zero. 

For our analysis of net charm fluctuations we use gauge field configurations
generated with the highly improved staggered quark (HISQ) action \cite{hisq}. Use of
the HISQ action in the charm sectors includes the so-called $\epsilon$-term and thus
makes our calculations free of tree-level order $(am_c)^4$ discretization errors
\cite{hisq}, where $m_c$ is the bare charm quark mass in units of the lattice spacing.  
These dynamical (2+1)-flavor QCD calculations have been carried out with a
strange quark mass ($m_s$) that has been tuned to its physical value and light $(u,\
d)$ quarks with mass $m_l/m_s =1/20$. In the continuum limit, the latter corresponds
to a light pseudo-scalar mass of about 160~MeV. The charm quark sector is treated
within the quenched approximation, neglecting the effects of charm quark
loops. Within the temperature range relevant for the present study, the quenched
approximation for the charm quarks is very well justified. Various lattice QCD
calculations using dynamical charm have confirmed that contributions of dynamical
charm quarks to bulk thermodynamic quantities, including the gluonic part of the trace 
anomaly as well as the susceptibilities of light,
strange and charm quarks, remain negligible even up to temperatures as high as 300
MeV \cite{STOUT211,MILC-C}. We note that these quantities
directly probe the influence of virtual quark pairs on observables
calculated at a fixed value of the temperature. Unlike in these cases
there is no simple observable known that would allow us to directly
calculate the pressure at fixed temperature. This may be the reason for
differences seen in current calculations of the pressure 
\cite{STOUT211,MILC-C} using quenched or dynamical charm.
In this work, we only use observables that are of the former type and also
do not require any multiplicative or additive renormalization.

The line of constant physics for the charm quark has been determined at zero
temperature by calculating the spin-averaged charmonium mass \cite{Yu}, $\frac{1}{4} (
m_{\eta_c} + 3 m_{J/\psi})$. For this purpose we used gauge field configurations
generated by hotQCD on lattices of size $32^4$ and $32^3\cdot48$ in the range of
gauge couplings, $6.39\le \beta= 10/g^2 \le 7.15$ \cite{bazavov,hotQCDHRG}.  On
finite temperature lattices with  temporal extent $N_\tau=8$, this covers the
temperature range\footnote{At finite lattice spacing  $f_K$ has been used to set the
temperature scale \cite{hotQCDHRG}.}  $156.8~{\rm MeV}\le T \le 330.2~{\rm MeV}$.  On
these lattices and for the slightly larger-than-physical light quark mass value used
in our calculations the transition temperature is $158(3)$~MeV, i.e. about $4$~MeV
larger than the continuum extrapolated results at the physical values of the light
and strange quark masses \cite{bazavov}. We consider this difference of about 3\% as
the typical systematic error for all temperature values quoted for our analysis,
which is not extrapolated to the physical point in the continuum limit.
 
The line of constant physics for the charm quark sector is well parametrized by
\begin{equation}
m_ca = \frac{c_0 R(\beta)  + c_2 R^3(\beta)}{1+d_2 R^2(\beta) } \; ,
\end{equation}
with $R(\beta)$ denoting the two-loop $\beta$-function of massless 3-flavor QCD and
$c_0=56.0$, $c_2 = 1.16\cdot 10^6$, $d_2=8.67\cdot 10^3$. On this line the charm quark
mass varies by less than 5\%. The ratio of charm and strange quark masses, $m_c/m_s$,
varies by about 10\%, with $m_c/m_s=12.42$ at $\beta=6.39$ and $m_c/m_s=11.28$ at
$\beta=7.15$.  

For most of our calculations we use data sets on lattices of size $32^3\cdot 8$. A
subset of these configurations has already been used for the analysis of strangeness
fluctuations \cite{strange}. These data sets have been enlarged and now contain up to
16700 configurations at the lowest temperature, separated by 10 time units in rational hybrid Monte
Carlo updates. Some additional calculations have been performed on coarser $24^3\cdot
6$ lattices, with fixed $m_c/m_s=12$, in order to check cut-off effects also in the
charm quark sector.  We summarize the statistics exploited in this calculation in
Table.~\ref{tab:stat}.  We calculate all the moments of net charm fluctuations needed
to construct up to fourth order cumulants that correlate net-charm fluctuations with
net baryon number, electric charge and strangeness fluctuations. As the calculation
of charm fluctuations is fast we can afford to use on each gauge field configuration
up to 6000 Gaussian distributed random source vectors for the inversion of the
charmed fermion matrix. This leaves us with statistical errors that mainly arise from
fluctuations in the light and strange quark sectors where we have used 1500 random 
source vectors for the inversion of the corresponding fermion matrices.

\begin{table}
\begin{center}
\begin{tabular}{|c|r||c|r|}
\hline
\multicolumn{2}{|c|}{$N_\tau =8$} & \multicolumn{2}{|c|}{$N_\tau =6$} \\ 
\hline
T[MeV] & \# conf & T[MeV] & \# conf \\
\hline
156.8 & 16700 &  & ~ \\
162.0 & 9520 & 162.3  & 7820\\
165.9 & 9000 & 166.7 & 3590\\
168.6 & 6130 & 170.2 & 5140 \\
173.5 & 5510 &  & \\
178.3 & 5500 &  & \\
184.8 & 5730 &  & \\
189.6 & 4930 &  & \\
196.0 & 6000 &  & \\
207.3 & 1800 &  & \\
237.1 & 1600 &  & \\
273.9 & 1600 &  & \\
330.2 & 1600 &  &\\
\hline
\end{tabular}
\end{center}
\caption{Number of configurations analyzed at different values of the temperature and
on different size lattices.}
\label{tab:stat}
\end{table}

\section{Partial pressure of open charm hadrons from fluctuations and correlations}

Our analysis of higher order cumulants of net charm fluctuations and their
correlations with net baryon number, electric charge and strangeness, closely follows
the concepts developed for our analysis of strangeness fluctuations \cite{strange}.
The large charm quark mass, $m_c\gg T$, however leads to some
simplifications. First of all, for temperatures a few times the QCD transition
temperature, Boltzmann statistics is still a good approximation for a free charm
quark gas. In the high temperature phase we can thus compare our results with
cumulants derived from a free massive quark-antiquark gas in the Boltzmann
approximation,
\begin{equation}
\frac{P_{c,free}(m_c/T,\vec{\mu}/T)}{T^4} = 
{3\over {\pi^2}} \left(\frac{m_c}{T}\right)^2 K_2({{m_c/T}}) \cosh \left( 
\frac{\hat{\mu}_B}{3}
+ \frac{2}{3} \hat{\mu}_Q + \hat{\mu}_C \right) \ ,
\label{fmc}
\end{equation}
where we used explicitly the quantum numbers of charm quarks. Another simplification
occurs at low temperatures, where we expect a hadron resonance gas to
provide a good description of cumulants of net charge fluctuations. At these
temperatures, the pressure of the hadronic medium receives contributions from
different open charm mesons and baryons. Using the fact that these 
hadrons carry integer conserved charges for baryon number
($|B|\le 1$), electric charge ($|Q|\le 2$), strangeness ($|S|\le 2$) and charm
($|C|\le 3$),  we  can separate the total open charm
contribution to the pressure in terms of different mesonic ($M_C$) and baryonic
($B_{C,i}$ with $i\equiv |C|= 1,\ 2,\ 3$) sectors,
\begin{equation}
\frac{P_C(T,\vec{\mu})}{T^4} = M_C (T,\vec{\mu})+B_C (T,\vec{\mu}) 
= M_C (T,\vec{\mu})+\sum_{i=1}^3 B_{C,i}(T,\vec{\mu}) \; .
\label{C-pressure}
\end{equation}
In this work, we further motivate the decomposition of the open charm pressure in terms of partial
pressures in different electric charge and strangeness sectors. 
In such cases, we decompose the corresponding partial pressures as,
\begin{equation}
\frac{P_{C,X}(T,\vec{\mu})}{T^4} = 
M_{C,|X|=1}(T,\vec{\mu}) +B_{C,|X|=1}(T,\vec{\mu}) + 
B_{C,|X|=2}(T,\vec{\mu})\; ,\; X=Q,\ S \; .
\label{PCQ}
\end{equation}

Due to the large charm quark mass, the masses of charmed baryons with $|C|=2$ or
$3$ are substantially larger than those of the $|C|= 1$ hadrons; e.g. $\Delta  =
m_{C=2}-m_{C=1} \simeq 1.2$~GeV. Even at $T\simeq 200$~MeV, i.e. well beyond the
validity range of any HRG model, the contribution of a $|C|=2$ hadron to
$P_C(T,\vec{\mu})/T^4$ thus is suppressed by a factor $\exp (-\Delta/T) \simeq
10^{-3}$ relative to that of a corresponding $|C|=1$ hadron. The latter thus will
dominate the total partial charm pressure, $P_C(T,\vec{\mu})/T^4 \simeq M_C
(T,\vec{\mu})+ B_{C,1}(T,\vec{\mu})$.  Similarly the baryon contributions to the
charged and strange partial charm pressures will be dominated by $|C|=1$ baryons
only.

The dominance of the $|C|=1$ sector in all fluctuation observables involving open
charm hadrons is immediately apparent from the temperature dependence of second and
fourth order cumulants of net-charm fluctuations, $\chi_2^C$ and $\chi_4^C$, as well
as the correlations between moments of net baryon number and charm fluctuations
($BC$-correlations).  As long as the strong interaction medium can be described by a
gas of uncorrelated hadrons these observables have simple interpretations in terms of
partial pressure contributions $M_C$ and $B_{C,i}$ evaluated at $\vec{\mu}=0$, 
\begin{eqnarray}
\chi_n^C &=& M_C +B_{C,1} + 2^n B_{C,2} + 3^n B_{C,3}\simeq M_C+B_{C,1}\; , 
\nonumber  \\
\chi_{mn}^{BC} &=& B_{C,1} + 2^n B_{C,2} + 3^n B_{C,3} \simeq B_{C,1} \; ,
\label{chin}
\end{eqnarray}
where $n,\ m >0$ and $n$ or $n+m$ are even, respectively.  Here and in the following
we often omit the arguments of the functions $M_C(T,0)$, $B_{C,i}(T,0)$.

The quantity $(\chi_4^C-\chi_2^C)/12$ is an upper bound for the contribution to the 
pressure from the $|C|>1$ channels in the open charm sector.
For all temperature values analyzed by us, we find that this quantity is less than 0.2\% of $\chi_2^C$. In
fact, for temperatures $T\le 200$~MeV the difference vanishes within errors.  This
may easily be understood as this difference is only sensitive to contributions of
baryons with charm $|C|=2,\ 3$; i.e. $\chi_4^C-\chi_2^C = 12 B_{C,2} + 72 B_{C,3}$ in
a gas of uncorrelated hadrons.  We thus conclude that up to negligible corrections
all cumulants of net-charm fluctuations, $\chi_n^C$, with $n>0$ and even, directly
give the total open charm contribution to the pressure in an HRG, $P_C\equiv P_C(T,0)
\simeq \chi_2^C$.  Moreover, each of the off-diagonal $BC$-correlations,
$\chi_{nm}^{BC}$, with $n+m > 0$ and even, approximates well the partial pressure of
charmed baryons, $B_C\equiv B_C(T,0)\simeq \chi_{mn}^{BC}$.  In
Fig.~\ref{fig:BQC}~(right) we show lattice QCD data for $\chi_4^C/\chi_2^C$.  In the
crossover region this ratio is close to unity. This confirms that at low temperature
the charm fluctuations $\chi_2^C$ and $\chi_4^C$ indeed are equally good
representatives for the open charm partial pressure.

\section{Melting of open charm hadrons}

In order to determine the validity range of an uncorrelated hadron resonance gas model
description of the open charm sector of QCD, without using details of the open charm
hadron spectrum, we analyze ratios of cumulants of correlations between net
charm fluctuations and net-baryon number fluctuations ($BC$-correlations) as well as
cumulants of net charm fluctuations ($\chi_n^C$).

As motivated in the previous section, a consequence of the dominance of the $|C|=1$  
charmed baryon sector in thermodynamic considerations is that, to a good approximation, 
$BC$-correlations in the hadronic phase obey simple relations as,
\begin{equation}
\chi_{nm}^{BC} \simeq \chi_{11}^{BC} \;\; ,\;\ n+m > 2\; {\rm and~even}\ .
\label{BC}
\end{equation}

\begin{figure}[!th]
\begin{center}
\includegraphics[scale=0.52]{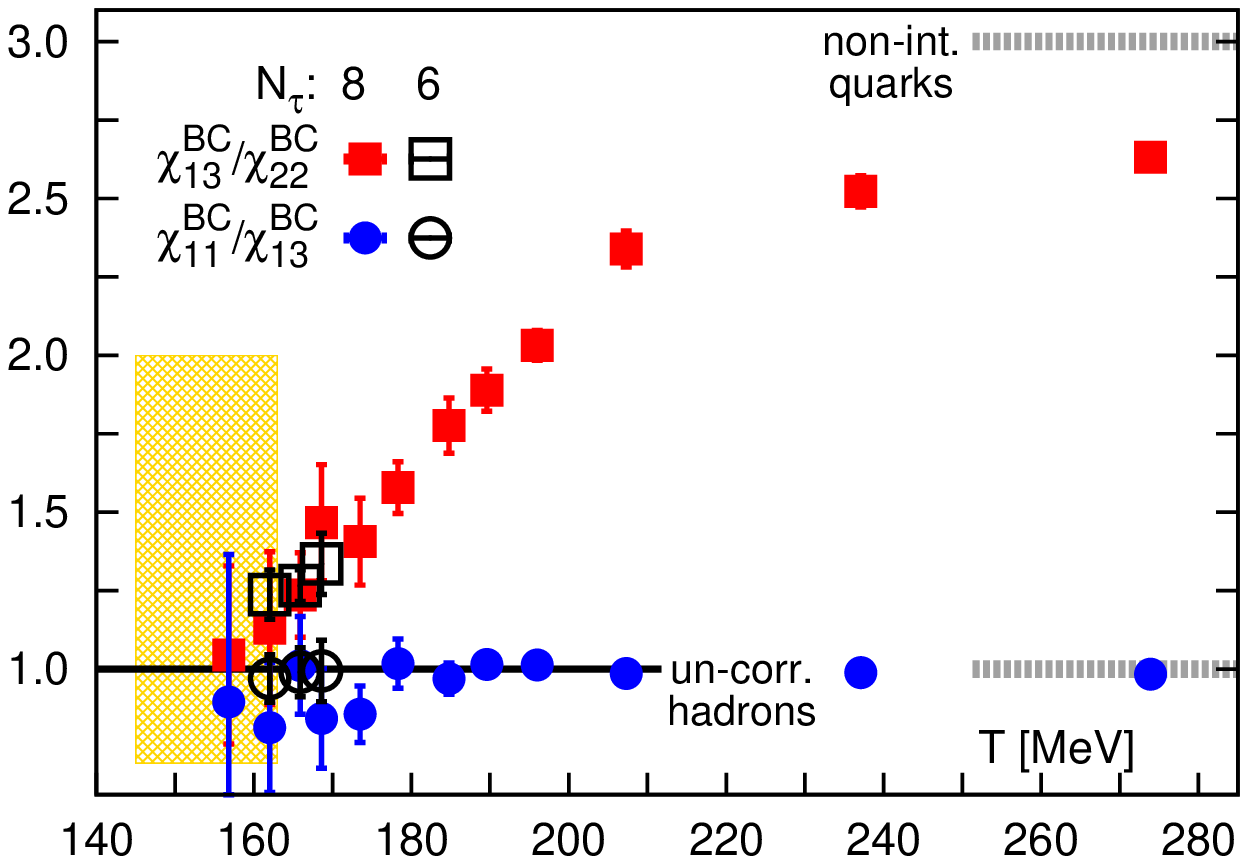}
\includegraphics[scale=0.52]{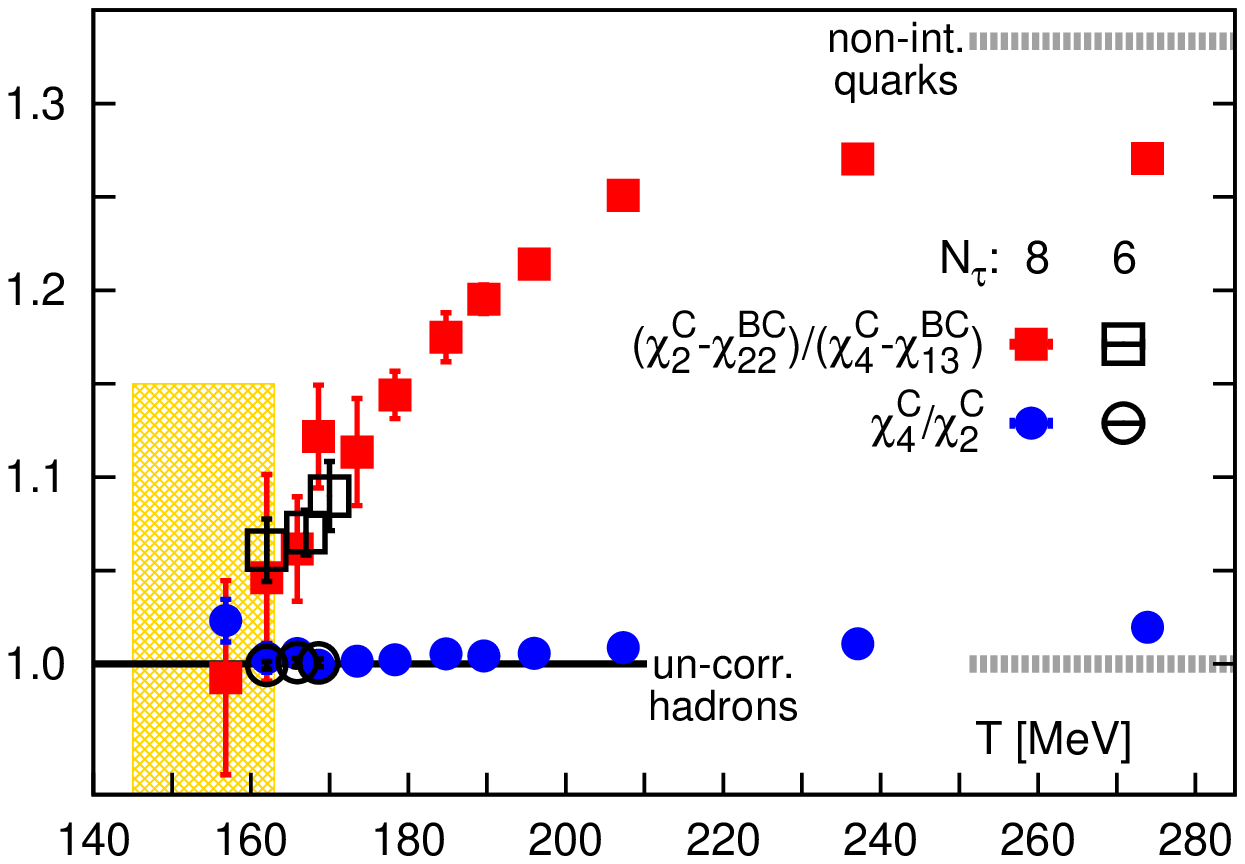}
\end{center}

\caption{The left hand figure shows two ratios of fourth order baryon-charm($BC$)
correlations. In an uncorrelated hadron gas both ratios receive contributions only
from charmed baryons. Similarly, for the right hand figure the ratio $\chi_4^C/\chi_2^C$ 
is dominated by and $(\chi_2^C-\chi_{22}^{BC})/(\chi_4^C-\chi_{13}^{BC})$  only
receives contributions from open charm mesons.  The horizontal lines on the right 
hand side of both figures show the infinite temperature non-interacting charm quark 
gas limits of the respective quantities. The shaded region indicates the chiral crossover 
temperature at the physical pion mass in the continuum limit, $T_c=(154\pm 9)$~MeV, determined
from the maximum of the chiral susceptibility \cite{bazavov}.  Calculations have been
performed on lattices of size $32^3\cdot 8$ (filled symbols) and $24^3\cdot 6$ (open
symbols). }

\label{fig:BQC}
\end{figure}

The ratio of any two of these susceptibilities, i.e. $\chi_{nm}^{BC}/\chi_{kl}^{BC}$
thus will be unity in a hadron resonance gas irrespective of its composition and the 
details of the baryon resonance spectrum. In Fig.~\ref{fig:BQC}~(left) we show the ratio
$\chi_{13}^{BC}/\chi_{22}^{BC}$. It clearly suggests that above the crossover region, 
an uncorrelated gas of charmed baryons does no longer
provide an appropriate description of the $BC$-correlations. Also shown in this
figure is the ratio $\chi_{11}^{BC}/\chi_{13}^{BC}$. It is consistent with unity for
all temperatures  because the relation $\chi_{1n}^{BC} = \chi_{11}^{BC}$ not only holds 
in a non-interacting charmed hadron gas (Eq.~\ref{BC}), but also is valid in an
uncorrelated charmed quark gas, as is easily seen from Eq.~\ref{fmc}.
Higher order derivatives with respect to baryon chemical potentials,
on the other hand, distinguish between the hadronic and partonic phases. 
E.g., one finds that for $n$ being odd, $\chi_{n1}^{BC} / \chi_{11}^{BC}=1$ in a 
hadron gas and $3^{1-n}$ in an uncorrelated charm quark gas.

Subtracting any of the $BC$-correlations from the quadratic or quartic charm
fluctuations provides an approximation for the open charm meson pressure in a gas of
uncorrelated hadrons. We thus expect for instance, the relation
\begin{equation}
M_C = \chi_4^C - \chi_{13}^{BC} =  \chi_2^C - \chi_{22}^{BC} \, .
\label{BC-meson}
\end{equation}
to hold at low temperatures.
Their ratio thus should be unity at low temperatures as long as the HRG description 
is valid. Fig.~\ref{fig:BQC}~(right) shows the ratio of the two observables introduced in
Eq.~\ref{BC-meson}.   It is obvious from the figure that also in the meson sector, 
an HRG model description breaks down in the crossover region at or close to $T_c$.

\begin{figure}[!th]
\begin{center}
\includegraphics[scale=0.52]{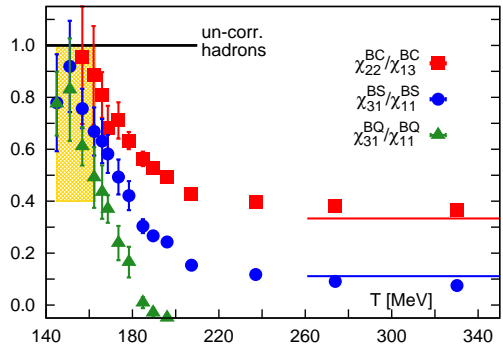}
\end{center}

\caption{Ratios of  baryon-electric charge($~BQ$), baryon-strangeness($~BS$)
and baryon-charm($~BC$) correlations calculated on lattices of size $32^3\cdot 8$. In
the case of $BQ$ and $BS$ correlations we show results from the (2+1)-flavor
calculations where $B$ and $Q$ do not contain any charm contribution. These data are
taken from Ref.~\cite{strange,freeze}. The shaded region shows the chiral crossover
region as in Fig.~\protect\ref{fig:BQC}. Horizontal lines on the right side show corresponding 
results for an uncorrelated quark gas. It should be noted that this limiting value is not defined for 
$\chi_{31}^{BQ}/\chi_{11}^{BQ}$ since the denominator as well as the numerator vanishes in 
perturbation theory up to ${\cal O}(g^4)$.}

\label{fig:BC_MC}
\end{figure}

The behavior seen in Fig.~\ref{fig:BQC} for correlations between net charm
fluctuations and net baryon number fluctuations, in fact, is quite similar to the
behavior seen in the strangeness sector ($BS$-correlations) \cite{strange} as well as
in the light quark sector which dominates the correlations between net electric
charge and net baryon number ($BQ$-correlations) \cite{hotQCDHRG}.  In
Fig.~\ref{fig:BC_MC} we show a comparison of ratios of cumulants of such correlations.
For the $BS$ and $BQ$ correlations with the lighter quarks we have two additional data 
points below $156~$MeV. In the charm sector we choose a 
ratio of cumulants involving higher order derivatives in the charm
sector as correlations involving only first order derivatives 
have large statistical errors.
These ratios all should be unity in a gas of uncorrelated hadrons. It is apparent
from Fig.~\ref{fig:BC_MC} that such a description breaks down for charge correlations
involving light, strange, or charm quarks in or just above the chiral crossover
region.

\section{Abundance of open charm hadrons}

\begin{figure}[!th]
\begin{center}
\includegraphics[scale=0.6]{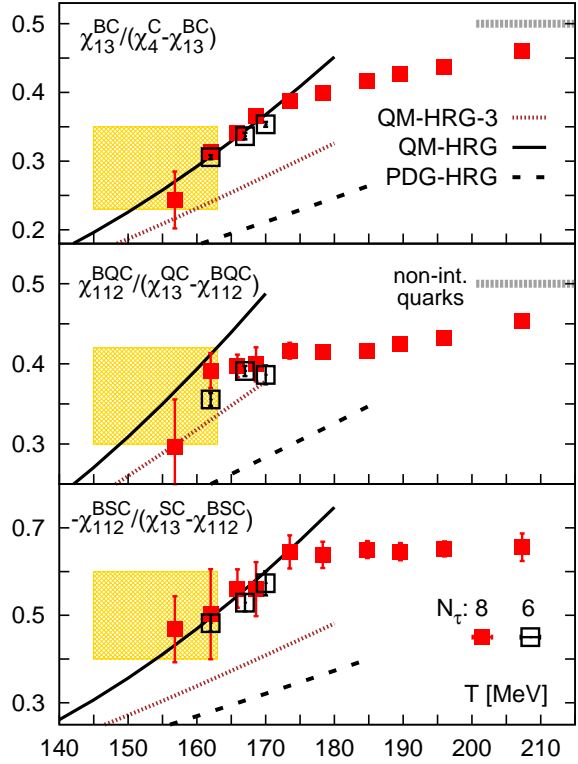}
\end{center}

\caption{Thermodynamic contributions of all charmed baryons, $R_{13}^{BC}$ (top), all
charged charmed baryons, $R_{13}^{QC}$ (middle) and all strange charmed baryons,
$R_{13}^{SC}$ (bottom) relative to that of corresponding charmed mesons (see \eq{ratios}). The
dashed lines (PDG-HRG) are predictions for an uncorrelated hadron gas using only the
PDG states. The solid lines (QM-HRG) are similar HRG predictions including also the
states predicted by the quark model of Ref.\ \cite{Ebertm,Ebert}. The dotted lines
(QM-HRG-3) are the same QM predictions, but only including states having masses $<3$
GeV.  The shaded region shows the QCD crossover region as in Fig.~\ref{fig:BQC}. 
The horizontal lines on the right hand side denote the infinite temperature non-interacting charm quark gas
limits for the respective quantities. The lattice QCD data have been obtained on
lattices of size $32^3\cdot8$ (filled symbols) and $24^3\cdot6$ (open symbols).  }

\label{fig:SC_QC}
\end{figure}

We now turn to the analysis of ratios of charge correlations and fluctuations that
are, in contrast to the ratios shown in Fig.~\ref{fig:BQC},  sensitive to some
details of the open charm hadron spectrum. 
We construct partial pressure components for the electrically charged charmed mesons 
and the strange-charm mesons, $M_{QC}\simeq\chi_{13}^{QC}-\chi_{112}^{BQC}$ and
$M_{SC}\simeq\chi_{13}^{SC}-\chi_{112}^{BSC}$, respectively. We also consider the partial 
pressure of all open charm mesons $M_C = \chi_4^C - \chi_{13}^{BC}$ as motivated in 
Eq.~\ref{BC-meson}. Using these observables we construct ratios with cumulants, which 
in an HRG receive contributions only from different charmed baryon sectors in the numerator,
\begin{equation} 
R_{13}^{BC} = \frac{\chi_{13}^{BC}}{M_C} \;\; ,\;\; 
R_{13}^{QC} = \frac{\chi_{112}^{BQC}}{M_{QC}} \;\; ,\;\;
R_{13}^{SC} = - \frac{\chi_{112}^{BSC}}{M_{SC}} \;\; .
\label{ratios}
\end{equation}
In an HRG, the first ratio just gives the ratio of charmed baryon and meson
pressure, $\left(R_{13}^{BC}\right)_{HRG} = B_C/M_C$. In the two other cases, the
numerator is a weighted sum of partial charmed baryon pressures in charge sectors $|X|=1$ and
$|X|=2$ with $X=Q$ and $S$, respectively.  These ratios are shown in
Fig.~\ref{fig:SC_QC}.

HRG model predictions for these ratios strongly depend on the relative abundance of the
charmed baryons over open charm mesons.  Shown in Fig.~\ref{fig:SC_QC} are results
obtained from the PDG-HRG calculation (dashed lines) and the QM-HRG (solid lines). 
Clearly in the temperature range of the QCD crossover transition, the lattice QCD data for these
ratios are much above the PDG-HRG model results. In all the cases, the deviation from the
PDG-HRG at $T=160$~MeV is 40\% or larger.  As discussed in Sec.~2, this may not be too
surprising as only a few charmed baryons have so far been listed in the particle
data tables. The lattice QCD results instead show good agreement with an HRG
constructed from open charm meson and baryon spectra calculated in a relativistic
quark model \cite{Ebertm,Ebert}. The difference in PDG-HRG and QM-HRG model
calculations mainly arises from the baryon sector (see Fig.~\ref{fig:hadronsPDG}).
The observables shown in Fig.~\ref{fig:SC_QC} thus provide first-principles evidence for a substantial
contribution of experimentally so far unobserved charmed baryons to the pressure
of a hadron resonance gas\footnote{It should be obvious that this contribution to the
pressure nonetheless is strongly suppressed relative to the contribution of the
non-charmed sector in  HRG models.}.  This is also consistent with a large set
of additional charmed baryon resonances that are predicted in lattice QCD 
calculations \cite{Edwards}.

\section{Conclusions}

We have calculated second and fourth order cumulants of net charm fluctuations and
their correlations with fluctuations of other conserved charges, i.e. baryon number,
electric charge and strangeness. Ratios of such cumulants indicate that a
description of the thermodynamics of open charm degrees of freedom in terms of an
uncorrelated charmed hadron gas is valid only up to temperatures close to the chiral
crossover transition temperature. This suggests that open charm hadrons start to
dissolve already close to the chiral crossover. Moreover, observables
that are sensitive to the ratio of the partial open charm meson and baryon pressures
as well as their counterparts in the electrically charged charm sector and the
strange-charm sector suggest that a large number of so far experimentally not
measured open charm hadrons will contribute to bulk thermodynamics close to the
melting temperature. This should be taken into account when analyzing the
hadronization of charmed hadrons in heavy ion collision experiments.

So far our analysis has been performed by treating the charm quark sector in quenched
approximation using fully dynamical (2+1)-flavor gauge field configurations as
thermal heat bath. This, in fact, seems to be appropriate for the situation met in
heavy ion collisions, where charm quarks are not generated thermally but are embedded
into the thermal heat bath of light and strange quarks through hard collisions at
early stages of the collision. We also do not expect that the cumulant ratios
analyzed here will change significantly by treating also the charm sector
dynamically. This, however, should be verified in future calculations.


\section*{Acknowledgments}
\noindent
This work has been supported in part through contract DE-AC02-98CH10886 with the U.S.
Department of Energy, through Scientific Discovery through Advanced Computing
(SciDAC) program funded by U.S. Department of Energy, Office of Science, Advanced
Scientific Computing Research and Nuclear Physics, the BMBF under grant 05P12PBCTA,
the DFG under grant GRK 881, EU under grant 283286 and the GSI BILAER grant.
Numerical calculations have been performed using GPU-clusters at JLab, Bielefeld
University, Paderborn University, and Indiana University. We acknowledge the support
of Nvidia through the CUDA research center at Bielefeld University.


\end{document}